\documentclass[11pt,journal]{IEEEtran}

\usepackage{graphicx}
\usepackage{epstopdf}
\usepackage{amsmath,amssymb,amsthm,mathrsfs,amsfonts,dsfont}
\usepackage{epsfig}
\usepackage{color}
\usepackage{subfigure}
\usepackage{cite}
\usepackage{diagbox}
\usepackage{multirow}
\usepackage{float}
\usepackage{stfloats}
\usepackage{hyperref}

\graphicspath{{figures/}}

\begin{document}
\title{Resource Rationing for Wireless Federated Learning: Concept, Benefits, and Challenges}
\author{Cong Shen, Jie Xu, Sihui Zheng, and Xiang Chen
\thanks{C. Shen is with the Charles L. Brown Department of Electrical and Computer Engineering, University of Virginia, Charlottesville, VA 22904, USA.}
\thanks{J. Xu is with Department of Electrical and Computer Engineering, University of Miami, Coral Gables, FL 33124, USA.}
\thanks{S. Zheng and X. Chen are with School of Electronics and Information Technology, Sun Yat-Sen University, China.}
}

\maketitle

\begin{abstract}

We advocate a new resource allocation framework, which we term \textit{resource rationing}, for wireless federated learning (FL). Unlike existing resource allocation methods for FL, resource rationing focuses on balancing resources across learning rounds so that their collective impact on the federated learning performance is explicitly captured. This new framework can be integrated seamlessly with existing resource allocation schemes to optimize the convergence of FL. In particular, a novel ``later-is-better'' principle is at the front and center of resource rationing, which is validated empirically in several instances of wireless FL. We also point out technical challenges and research opportunities that are worth pursuing. Resource rationing highlights the benefits of treating the emerging FL as a new class of \textit{service} that has its own characteristics, and designing communication algorithms for this particular service.

\end{abstract}

\section{Introduction}
\label{sec:intro}

Federated Learning (FL) is an emerging distributed machine learning (ML) paradigm that has many attractive properties. In particular, FL caters to the growing trend that massive amount of the real-world data is generated at the edge devices, and the combination of growing storage and computational power of devices and the increasing concern over transmitting private information to a central server has made it attractive to store data and train ML models locally on each device. The power of FL has been realized in commercial devices (e.g.,  Pixel 2 uses FL to train ML models to personalize user experience) and ML tasks (e.g., Gboard uses FL for keyboard prediction) \cite{bonawitz2019towards}.

Despite being recognized as one of the {primary bottlenecks} of FL since its inception \cite{mcmahan2017fl}, research on the communication aspect in the FL pipeline has not been on par with the learning component. Early research on communication-efficient FL largely focused on reducing the amount of information to be transmitted, and did not touch the actual communication algorithm and protocol design \cite{konecny2016fl,sattler2019robust}.  More recent research started to fill this void from a wireless communication and networking point of view \cite{wang2019adaptive,yang2019scheduling,zhu2019broadband,amiri2020federated,yang2020federated,xu2020client,shen2020gc,wei2021}.  In general, the principle is to balance learning performance and communication efficiency via, e.g., device selection, bandwidth allocation, and power control. It has been shown that combining (communication-oriented) adaptive resource allocation with (ML-oriented) efficient model representation leads to a better overall implementation, in particular for {wireless federated learning} which is envisioned to be among the mainstream deployment scenarios of edge ML \cite{zhu2020,lim2020federated}.

While the early results demonstrate the potential of jointly optimizing communication and computation for wireless FL, {the communication design has not been tailored to the unique characteristics of FL}. In particular, an intrinsic and fundamental property of FL has largely been ignored:  FL is a long-term process consisting of many progressive learning rounds that \textit{collectively} determine the learning performance. Because of this progressive nature, learning rounds may have varying significance towards the convergence rate and final model accuracy, and thus should weigh differently when allocating communication resources. However, almost all existing works treat every learning round as equally important and perform resource allocation independently across learning rounds. Specifically, the common underlying assumptions made in these works \cite{wang2019adaptive,yang2019scheduling,zhu2019broadband,amiri2020federated,yang2020federated} include that, in every learning round, the same total bandwidth is available, the same (exact or average) number of mobile devices are selected, or the same energy constraints are imposed. Resource allocation is then performed within the round under these constraints. While this view of static resource allocation across time simplifies the problem, it may lead to inefficient utilization of the scarce communication resources and consequently degrade the FL performance.

We advocate a novel resource allocation framework for wireless FL, which we term \textit{resource rationing} to emphasize balancing resources over time so that the long-term impact on the FL performance is explicitly captured.  The fundamental principle that differentiates resource rationing from existing wireless resource allocation is the focus on allocating different resources across learning rounds (in addition to possible resource allocation within each round) to optimize the convergence of FL \cite{xu2020client,shen2020gc}.  This novel framework originates from a holistic view of the resource management problem in FL: in order to achieve the best possible learning outcome with fast convergence, one has to ``ration'' the limited resources, consuming little at the beginning and gradually increasing towards the end. We introduce the basic concept of resource rationing and show that the \textit{``later-is-better''} principle is general and applies to different resources in a wireless FL system.  We also discuss several technical challenges and research directions to advance resource rationing and wireless FL in general.

\section{Resource Rationing for Wireless FL}
\label{sec:concept}

\subsection{Motivation}

\begin{figure}
  \centering
  \includegraphics[width=0.99\linewidth]{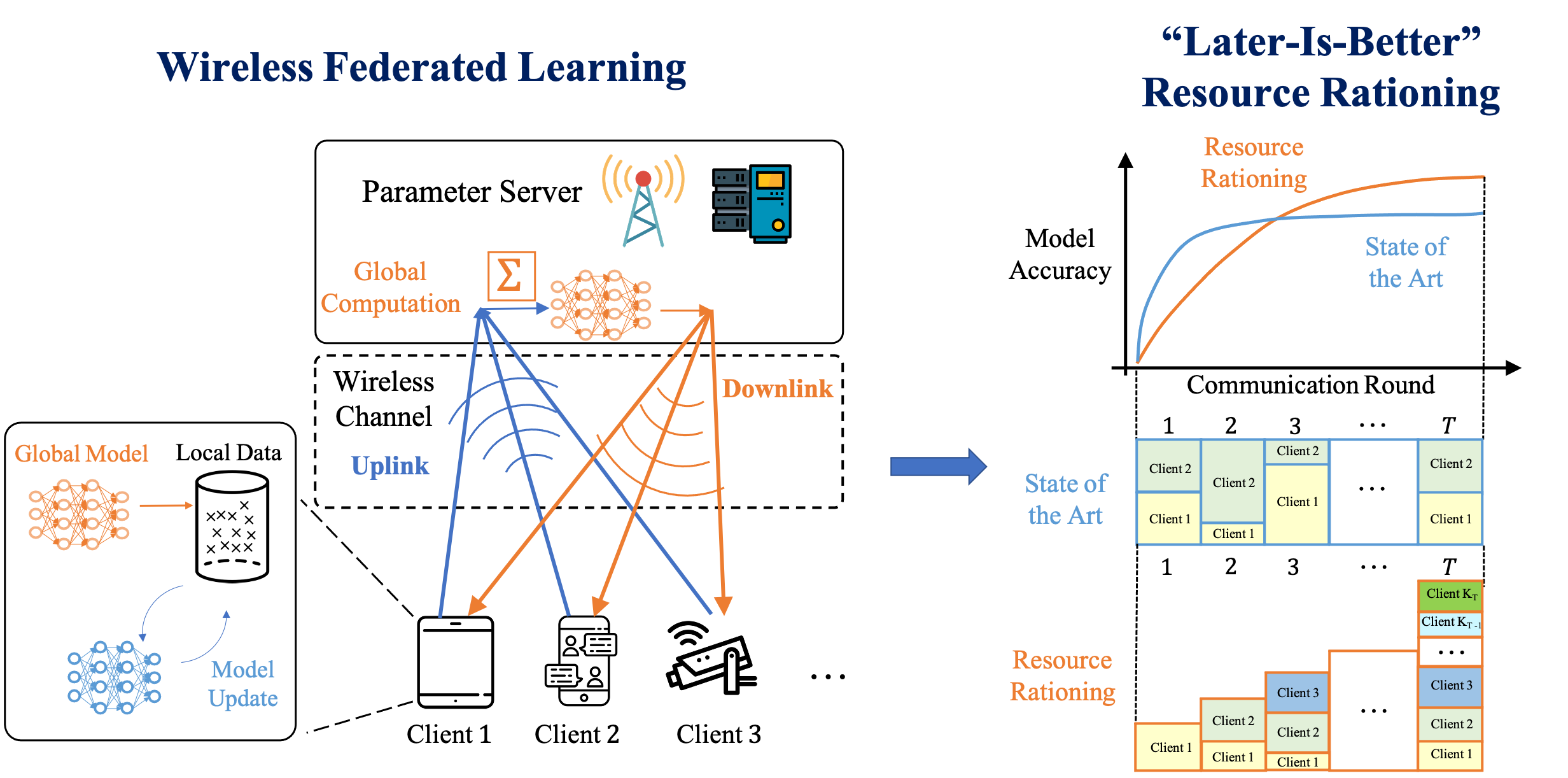}
  \caption{Overview of the proposed resource rationing framework in a wireless FL system. The ``later-is-better'' principle for resource rationing is illustrated with the example of clients rationing that is discussed in Section~\ref{sec:ben2}.}
  \label{fig:1}
\end{figure}

A representative wireless FL system is depicted in Fig.~\ref{fig:1}. It works by iteratively executing the following steps: (1) the parameter server broadcasts the current global ML model to participating mobile devices ({downlink communication}); (2) Starting from the received global model, each device trains a local model using its own dataset ({local computation}); (3) mobile devices upload their updated local  models to the parameter server ({uplink communication}); (4) the parameter server then aggregates these updates to generate a new global model ({global computation}). These four steps are collectively called a learning round. In this way, FL lets each mobile device evolve its own model using local data, while synchronizing the model training among different mobile devices via occasional model aggregation.

The primary drive for resource rationing is that FL is very different than most of the services for which current communication systems are designed. The purpose of communication in today's system is to deliver the information bits efficiently and reliably from one point to another. Although this is still true for FL, the ultimate goal of communication is to facilitate machine learning (e.g., collaboratively training a neural network to achieve the best classification accuracy), which has its own characteristics.  Thus, taking an isolated view of each individual communication phase and optimizing the resource allocation within each round, albeit still meaningful, misses the opportunity to allocate resource towards the ultimate prize -- enabling fast convergence of ML training to an accurate model.  This has motivated us to treat FL as a new class of service that has its own characteristics, and the problem we want to solve is how to {\em holistically} allocate the scarce wireless resource to optimize the particular service of FL.

\subsection{Impact of Resource}

\begin{figure}
  \centering
  \includegraphics[width=0.9\linewidth]{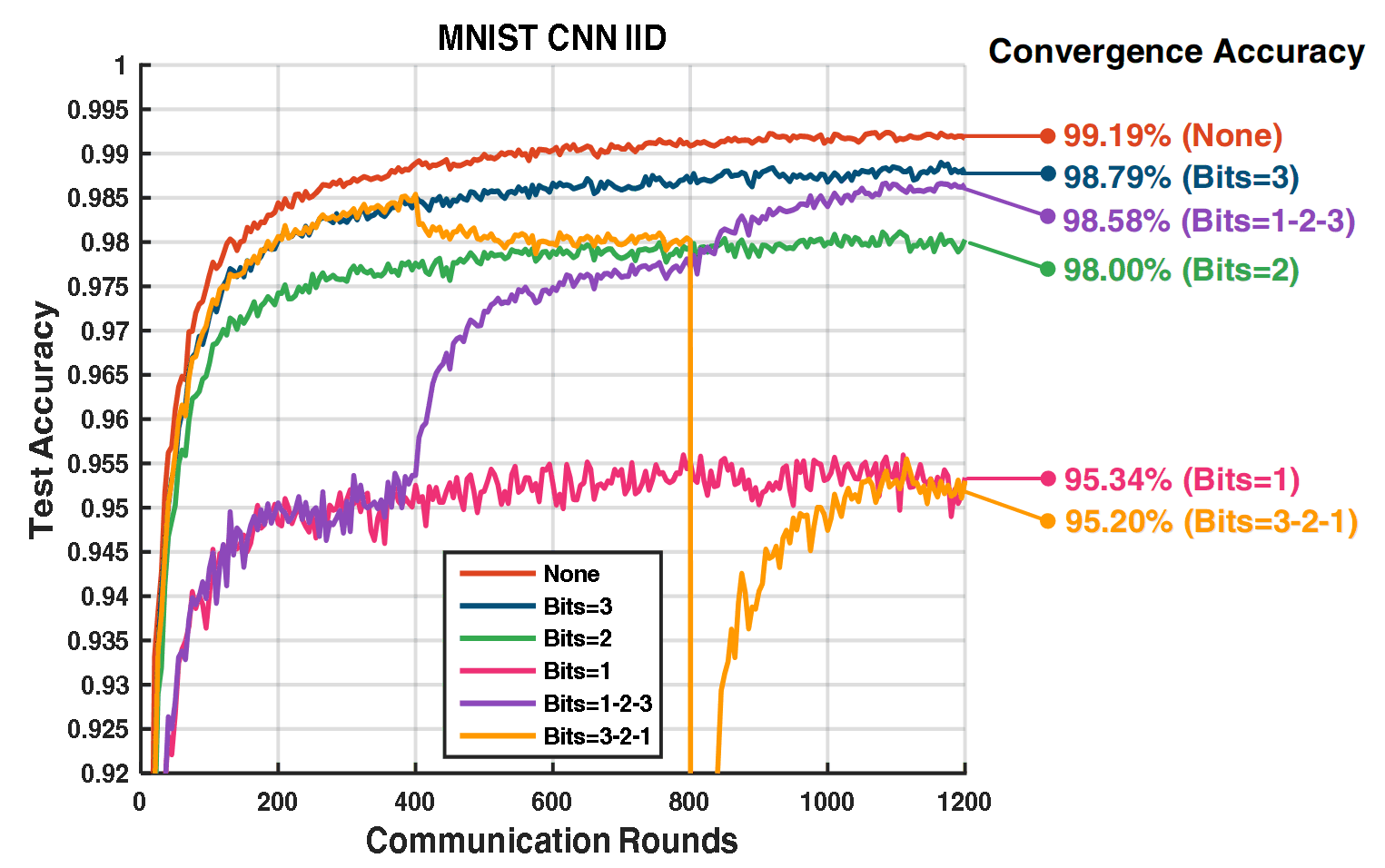}
 \caption{Bandwidth reduction and resource rationing on training a convolutional neural network (CNN) for the MNIST digit recognition task with IID local datasets. Tuned quantization and stochastic rounding are utilized. 10 out of a total of 2000 clients are uniformly randomly selected in each round. Each curve is obtained by averaging 10 independent runs of the FL process. The final model accuracy is averaged over the last 200 learning rounds for all methods.}
  \label{fig:2}
\end{figure}

The first step towards resource rationing is to understand the impact of resource to the overall learning accuracy and convergence. Taking communication bandwidth as an example. 
It is well known that the power of deep learning comes from the  ``depth'' of the network and, correspondingly, significantly many weight coefficients. For example, the original {ResNet-50} has over 23 million parameters. Even for the widely popular {MobileNet} which is specifically designed for devices with limited computing resource or limited power, the most economical version has 0.2 million parameters (\textit{0.25 MobileNet-128} model, face attribute classification) \cite{howard2017mobilenets}. A standard 32-bit floating-point representation of the updated weight coefficients leads to an uplink transmission of $6.4\times 10^6$ bits per user per round, which requires significant communication bandwidth. It is thus crucial to decide \textit{what} and \textit{how} to communicate for the latest model with limited communication resources.

One way to reduce the communication bandwidth is to compress the weights before each uplink and downlink communication round. Intuitively, the uplink communication is more resource constrained since the mobile devices are less powerful than the parameter server (e.g., base station). We thus focus on quantizing the locally updated weights and evaluating the performance impact to the overall FL convergence in a well adopted MNIST digit recognition task \cite{mcmahan2017fl}. Through a carefully designed quantization method \cite{shen2020gc} that adjusts the quantization gain based on the dynamic range of the weights, and the adoption of {stochastic rounding}, we are able to significantly reduce the communication bandwidth at negligible loss accuracy and convergence rate as shown in Fig.~\ref{fig:2}  -- this particular example shows that for the independent and identically distributed (IID) dataset, by using only 
$9.4\%$ bandwidth of the floating-point baseline, we are able to achieve 
$99.6\%$ of the baseline accuracy, at a convergence rate that remains almost the same. Similar observation can be made from different ML tasks and different configurations; see \cite{shen2020gc} for more details. This shows that it is possible to significantly reduce the communication resources while preserving the learning performance.

\subsection{``Later-Is-Better''}
\label{sec: earlylate}
With a better understanding of how much overall resource is needed, we now consider the general resource rationing framework that absorbs existing wireless-specific resource allocation and ``elevates'' the problem dimension to manage resources over learning rounds. The immediate question is whether a general principle exists for resource rationing, i.e., for a given total resource budget, what is the rule of thumb to ration resources?

\begin{figure}
  \centering
  \includegraphics[width=0.7\linewidth]{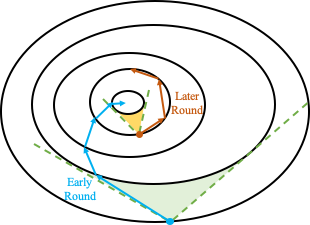}
  \caption{Early learning rounds have more leeway for gradient direction to be wrong than later rounds.}
  \label{fig:3}
\end{figure}

To gain some insight into this question, we need to build an understanding of how FL works across learning rounds. Predominantly, FL tasks involve training a deep neural network (DNN) with (mini-batch) stochastic gradient decent (SGD), which uses a smaller number of data samples to calculate an approximated gradient for updating the model parameters. When data is further distributed among multiple clients, FL lets each client perform SGD using their local data with occasional synchronization by model averaging. A careful examination of the weight update and model averaging mechanisms for FL, e.g., FedAvg \cite{mcmahan2017fl}, reveals an important feature that has not been incorporated in FL resource allocation: when the current weight is far from the optimal value, a rough gradient estimate is enough to find a descent direction.  As the weight starts approaching the optimal solution, however, noisy gradient estimates frequently fail to produce descent directions and do not reliably decrease the objective. An illustration of this critical hypothesis is given in Fig.~\ref{fig:3}, where early rounds enjoy much larger leeway (the large green area) in choosing gradient directions than later rounds (the small yellow area), where the same level of gradient noise may lead to deviation from convergence. This naturally leads to a ``later-is-better'' resource rationing rule: preserve resources at the early rounds of FL to exploit the tolerance of noisy gradient estimates, and spend the saved resources at later rounds to produce more accurate gradient estimate, thereby achieving an overall better learning performance.

\section{Benefits of Resource Rationing}
\label{sec:ben}

We instantiate resource rationing and demonstrate its benefits with three specific examples. On the physical layer, we study how to ration a given bandwidth budget over the entire learning period and how different allocations affect the final model accuracy and convergence rate. On the MAC layer, we study varying client selection strategies and evaluate how the convergence responds to client rationing. Lastly, we give an example of joint design that simultaneously rations clients selection and power control.

\subsection{Bandwidth Rationing}
\label{sec:ben1}

We have seen that for the specific experiment in Fig.~\ref{fig:2}, a 3-bit weight representation achieves near-optimal learning performance and clearly outperforms 2-bit and 1-bit representations, for a constant bandwidth allocation. We now elevate the problem setting and fix the total uplink communication bandwidth consumption across all learning rounds as $2T$ per weight per client, where $T$ is the total rounds of FL. Note that this would correspond to a 2-bit weight representation in the constant allocation, but we now  evaluate two different bit rationing schemes: smaller number of bits at the beginning and larger number of bits later, and vice versa.  The results are also shown in Fig.~\ref{fig:2}, where the increasing pattern uses 1-bit weight representation for the first one third of the rounds, 2-bit for the middle one third, and 3-bit for the final one third. The decreasing pattern is the exact reverse of the increasing pattern.  The results suggest that ``later-is-better'' indeed achieves much improved performance: with the same total bandwidth as the constant 2-bit representation, it achieves a final learning accuracy of the 3-bit representation. However, this model accuracy improvement comes at the cost of reduced convergence rate: since early rounds use less bandwidth, the initial convergence is rather slow, which is predicted by the SGD analysis in Section~\ref{sec: earlylate}. This phenomenon reveals a fundamental tradeoff that may exist between model accuracy and convergence rate, which is worth further investigation. We also see that the decreasing patter behaves poorly, as it starts out strong but ``starves'' at the end, converging to the 1-bit weight performance.

\subsection{Clients Rationing}
\label{sec:ben2}

\begin{figure}
  \centering
  \includegraphics[width=0.9\linewidth]{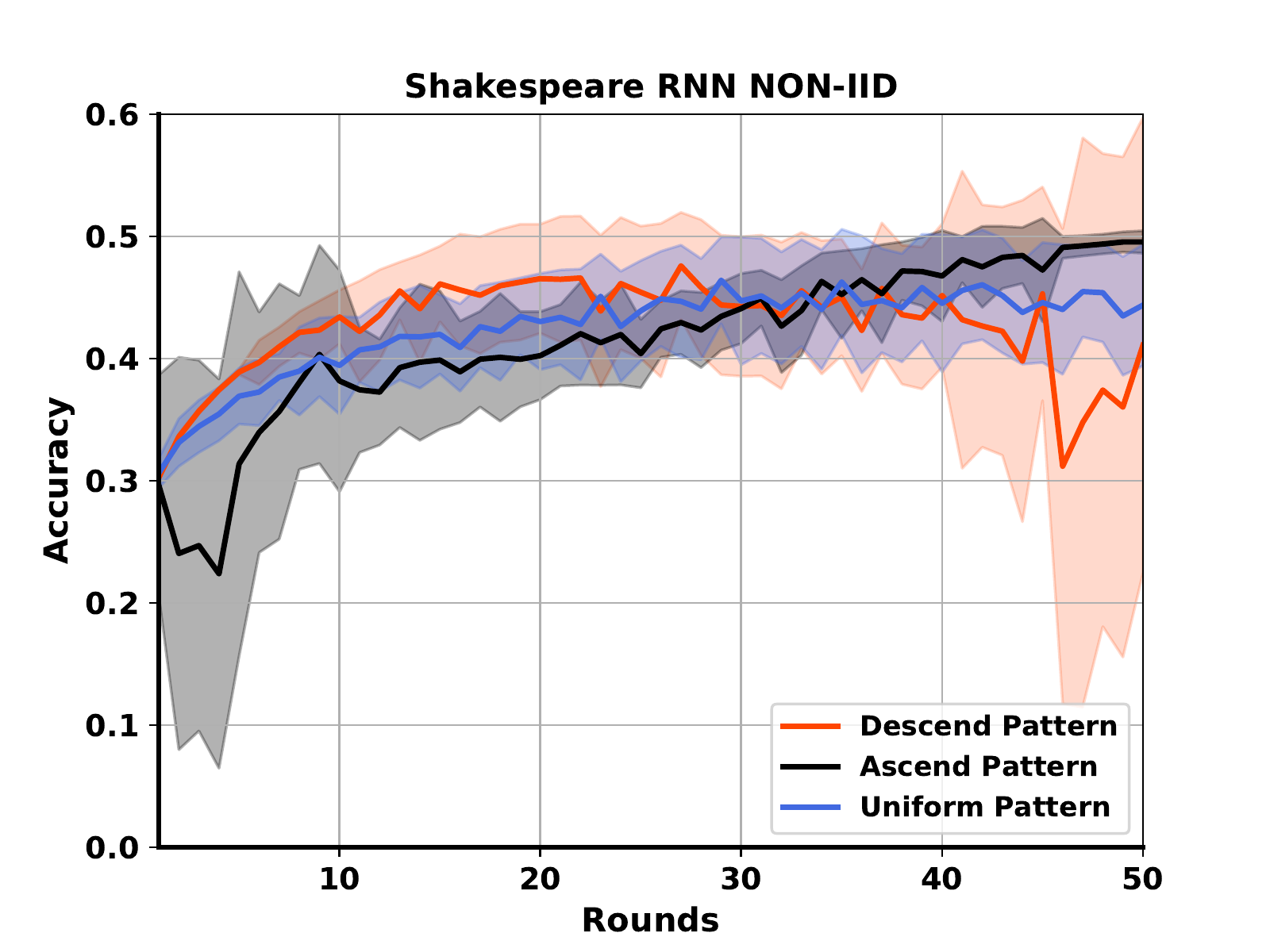}
  \caption{Convergence versus the communication rounds with varying client selection patterns. A pre-trained RNN that generates ASCII characters is refined via FL on the Shakespeare dataset, which is divided into 715 non-IID local datasets. A certain number of clients are uniformly randomly selected in each round according to the selection pattern. Each curve is obtained by averaging 30 independent runs of the FL process.}
  \label{fig:4}
\end{figure}

Similarly, one can also take a long-term perspective and study how different temporal client selection patterns lead to different learning performances.  The experiment is on another standard FL task -- train a recurrent neural network (RNN) for text generation on the Shakespeare dataset with FedAvg \cite{mcmahan2017fl}. Keeping the total number of participating clients throughout the entire FL process constant, we evaluate three patterns as follows. The state of the art corresponds to the \textit{Uniform} selection -- in each round, 5 clients are randomly selected. We consider two other rules: (1) \textit{Ascend} -- the number of randomly selected clients linearly increases from 0 to 10; (2) \textit{Descend} -- the number of randomly selected clients linearly decreases from 10 to 0.  Fig. \ref{fig:4} reports the convergence results of these three methods, which clearly show that selecting more clients in later FL rounds not only results in much higher model accuracy than selecting more clients in earlier FL rounds, but also is much more robust -- its standard deviation (shown as the green shaded area) at the end of training is significantly smaller than others (shown as the blue and orange shaded areas).

\subsection{Joint Design}
\label{sec:ben3}

\begin{figure}
  \centering
  \includegraphics[width=0.9\linewidth]{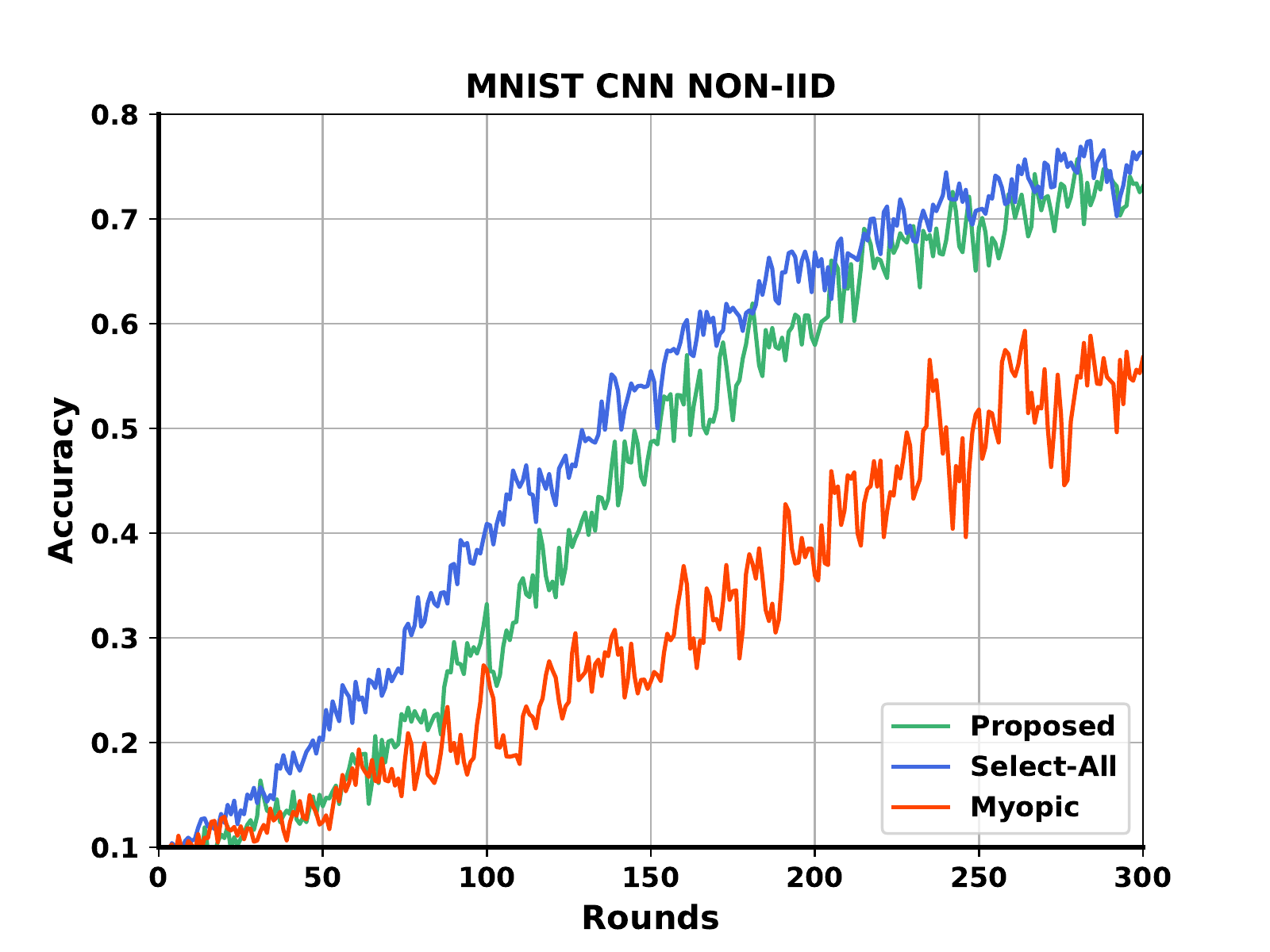}
  \caption{Convergence versus the communication rounds with a joint resource rationing design. A CNN for digit recognition is trained on the MNIST dataset, which is divided into 3383 local non-IID datasets. In every round, 10 clients are available and participating clients are selected among these clients. Each curve is obtained by averaging 30 independent runs of the FL process.}
  \label{fig:joint}
\end{figure}
A joint design of resource rationing among multiple types of resource can also be done to improve the final FL performance \cite{xu2020client}. Let us consider again the MINST digit recognition task, but the local datasets are non-IID. Three different client selection methods are investigated. The \textit{Select-All} method represents the ideal case, which selects all clients in every round and ignores the energy constraints; the \textit{Myopic} method imposes the same energy constraint in every round and selects as many clients as possible under this constraint, representing the state-of-the-art; the \textit{Proposed} method adaptively rations energy resources by performing joint client selection, bandwidth allocation and power control given the current wireless channel conditions, following the ``later-is-better'' principle (see \cite{xu2020client} for the detailed design). The convergence results of these methods are shown in Fig. \ref{fig:joint}. The eventual FL accuracy by using the proposed method far outperforms \textit{Myopic} and is close to the ideal case \textit{Select-All}, despite significantly reduced energy consumption (roughly $50\%$).

\section{Challenges and Opportunities}
\label{sec:chall}

Through the understanding of FL convergence and specific resource rationing schemes of bandwidth and clients allocation, we have established a general ``later-is-better'' resource rationing principle. This is a promising framework that intimately connects communication to FL.  In the following, we highlight some challenges associated with advancing this novel paradigm, and present research opportunities that we believe are worth pursuing.

\subsection{Theoretical Foundation}

Resource rationing for wireless FL must rely on a rigorous analysis of the varying significance of different learning rounds, and a deeper understanding of how resource rationing across time influences the convergence of FL. We already had a glimpse, through the bandwidth allocation example, that there may exist a fundamental tradeoff between model accuracy and convergence rate with a given total budget. However, despite the significant effort in establishing the convergence behavior of different FL algorithms under various regularity conditions, there is no research to directly connect FL convergence to the varying resource at each learning round. This theoretical foundation is difficult to establish but, unfortunately, is absolutely critical to enable a \textit{principled} design for resource rationing algorithms with proven performance guarantees.

\subsection{Temporal Variation in Wireless Systems}

\begin{figure}
  \centering
  \includegraphics[width=0.9\linewidth]{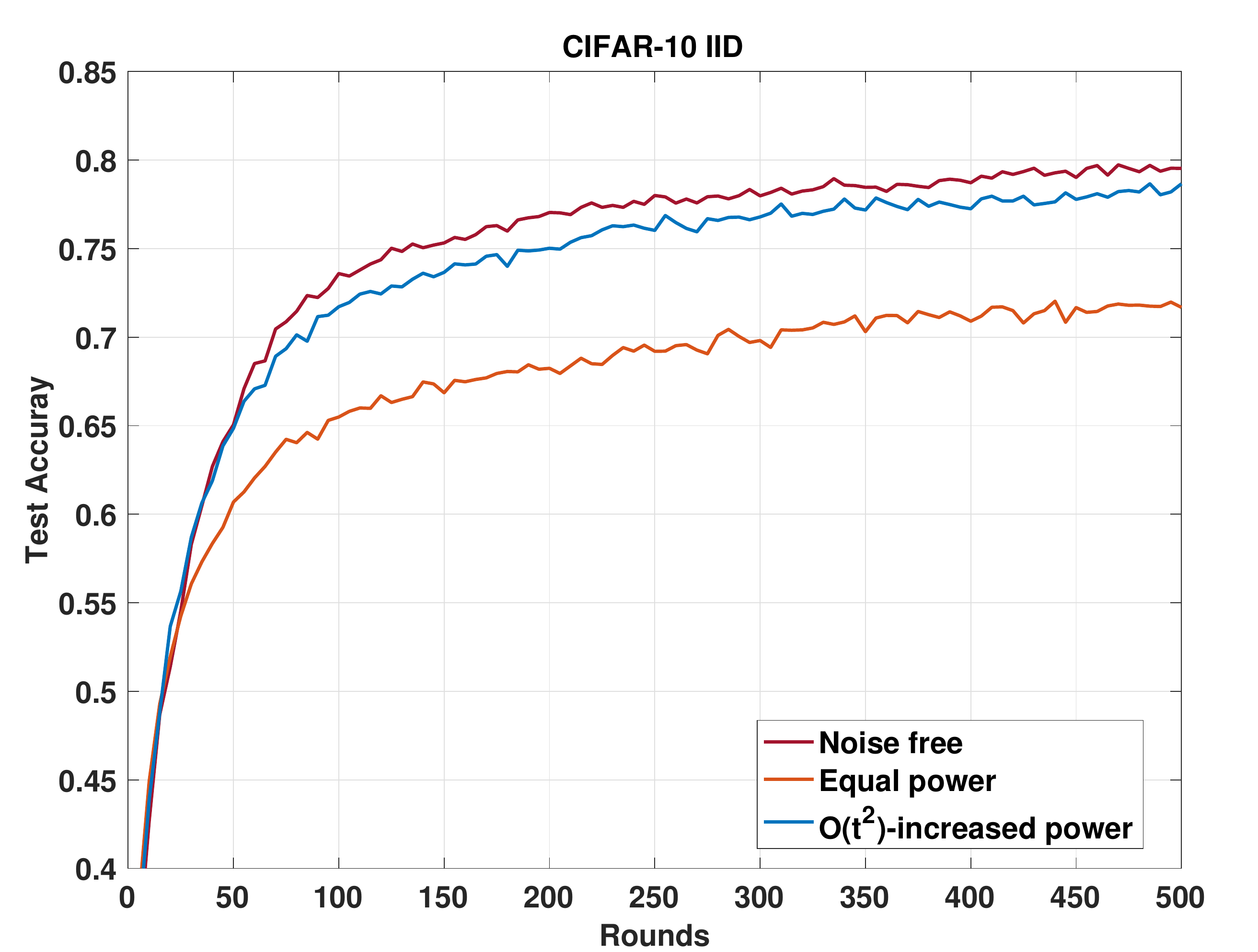}
  \caption{Comparing the performance of transmit power control \cite{wei2021} to the baselines with partial clients participation, model transmission and IID dataset for FL on the CIFAR-10 dataset. A single-cell multi-user cellular system with broadband analog aggregation in \cite{zhu2019broadband} for \textsc{FedAvg} is simulated where user devices participate in FL over wireless uplink and downlink communications. The results are averaged over 10 independent runs.}
  \label{fig:gtb}
\end{figure}

Wireless channels are dynamic and unpredictable in nature. When wireless channel characteristics are incorporated, {\em causality} issues may arise, making temporal resource rationing and applying the ``later-is-better'' principle challenging. For example, uploading the same updated model incurs different energy consumption under different wireless channel conditions.  The problem becomes further complicated because FL is a multi-user system where mobile devices are heterogeneous in terms of the experienced wireless environments, computing capabilities and resource constraints.  The proposed resource rationing framework, however, operates on the time scale of communication rounds, which allows for the flexibility to incorporate existing or future ``fast'' resource allocation mechanisms or prediction methods to handle temporal variations. We illustrate this flexibility using transmit power control as a use case \cite{wei2021}, and Fig.~\ref{fig:gtb} illustrates the performance advantage of combining the ``later-is-better'' principle of resource rationing with an inner-loop power control that handles channel fading and interference. Note that both ``equal power'' and ``$O(t^2)$-increased power'' consume the same total energy, and both implement the power control method proposed in \cite{zhu2019broadband} to enable analog aggregation in each round. Clearly, by deploying the resource rationing principle on top of the existing power control method, we can further improve the learning performance to be very close to the noise free benchmark, which has perfect communications. This also leads to several interesting future research directions, such as jointly designing resource rationing with temporal prediction (possibly leveraging ML), and performing real-world experiments to validate and evaluate the developed resource rationing framework.

\subsection{Generalization and Extension}

The concept of resource rationing is general and can be applied to a broad spectrum of resources whenever there is the flexibility to dynamically allocate them across learning rounds. In the physical layer, coding rate and modulation are another type of resource where we can attempt to develop novel adaptive coding and modulation methods to realize resource rationing.  In addition, from a pure learning perspective, there is no difference whether the updated model or its other forms are communicated between the clients and the server. However, this choice would affect the communication efficiency. For example, transmitting only the model difference as opposed to the updated model itself reduces the dynamic range. How to combine this feature with bandwidth rationing is an interesting research problem \cite{shen2020gc}. At the same time, the model difference becomes more sparse as the global model gradually converges. How to leverage the \textit{sparsity} in model difference for the communication design is another interesting problem. At the MAC layer, how to apply resource rationing to client selection, bandwidth allocation, power control, or a combination of them to accelerate learning convergence and model accuracy is worth exploring. Cross-layer designs can also be considered, e.g., one may be able to trade off coding and modulation for participating clients.  These different aspects collectively form the backbone of resource rationing in FL, and will constitute a major technological breakthrough that advances future applications.

\subsection{Complexity and Scalability}
The value of FL increases with more clients participating in the system, but the problem complexity of resource rationing will also increase considerably. Taking client selection in FL as an example, the selection space size is combinatorial with respect to the total number of clients. When there are many clients, searching for the optimal solution can be very difficult. To enable fast and effective resource rationing in large-scale wireless FL networks, designing low-complexity and/or distributed algorithms is essential. While pure optimization-based algorithms may still be worth exploring, it is interesting to leverage the generalization power of machine learning to develop a joint optimization and learning approach. A machine learning model may be trained on past resource rationing decisions and one can use this ML model to adjust future resource rationing.

\subsection{Beyond Communication Resources}

The resource rationing principle can be extended beyond allocating communication resources. For example, to guarantee learning convergence, it is required in SGD that the stepsize vanishes as time evolves. Because of the decaying stepsize, learning in later rounds is forced to be slow. In addition, finding the optimal learning rate schedule also requires an expensive grid search over all possible parameter values. Can we avoid using a pre-determined decaying stepsize schedule but rather automatically adapt the stepsize in each learning round? We may attempt to apply the resource rationing principle to reduce the stepsize by evaluating its impact to the convergence rate.

\section{Conclusions}
\label{sec:conc}

In this article, we have argued for a new resource rationing framework for wireless federated learning. Resource rationing takes a holistic view of the resource allocation problem and attempts to balance the resource consumption across the entire learning period, with the goal of maximizing the final ML model accuracy and convergence rate. This intuition has led to an interesting ``later-is-better'' principle, where we have demonstrated with several examples that reserving resources at the beginning and spending them later is beneficial to the performance of FL. A theoretical intuition was also provided based on stochastic gradient descent. Future directions and challenges were presented to spark research activities. Philosophically, resource rationing represents an example of tailoring communication to the characteristics of FL, and other components of the communication system for wireless FL may similarly benefit from this holistic view.

\bibliographystyle{IEEEtran}
\bibliography{FedLearn2}

\end{document}